*PREPRINT*

# GAMIFICATION IN THE AGE OF AI: SURVEYING PLAYER PERCEPTIONS OF MOTIVATION, MANIPULATION, AND DATA PROTECTION

*Kurt Horvath*
*Institute of Information Technology*
*University of Klagenfurt*
*Austria*

*Tom Tucek*
*Institute of Information Technology*
*University of Klagenfurt*
*Austria*

**ABSTRACT**

Modern games and applications increasingly employ artificial intelligence (AI), user modeling, and gamification to influence user behavior, sustain engagement, and monetize attention. Although these mechanisms can enhance motivation and enjoyment, they also raise concerns about manipulation, personal data usage, and transparency. To examine how players perceive these dynamics, we conducted a survey with 28 adult participants. Respondents expressed strong concerns about the collection and exploitation of personal information in games and gamified applications, while also reporting low awareness and a limited sense of control over how their data is used. They evaluated progression features such as leveling systems and leaderboards as motivating. In contrast, they rated retention-focused mechanics such as daily login rewards and random reward reveals as highly manipulative. Participants reported that these systems can create fear of missing out or resemble gambling. Participants also expressed support for regulatory measures that limit exploitative design and protect vulnerable users. These findings reveal a clear gap between increasingly powerful engagement strategies and user trust, illustrating the need for more responsible and transparent gamification practices.

## 1. INTRODUCTION

Video games have long driven advances in hardware and software innovation (Sousa et al. (2022); Wang et al. (2024)). Networked play transformed gaming from local interaction to a connected, social experience. Today, platforms such as handheld 5G phones and standalone VR headsets reflect gaming's shift toward pervasive, network-native engagement. Meanwhile, gamification extends game-like mechanics into everyday activities (Kim (2015)), shaping motivation in education, health, and commerce. IoT devices and AI systems increasingly personalize gameplay by sensing context and orchestrating computation (Kimovski et al. (2021)). While these advances improve responsiveness and user experience, they also raise concerns about personal data collection and behavioral influence (Caponetto et al. (2014); Hamari et al. (2014); Nyström (2021); Sardi et al. (2017); Zagal et al. (2013)). As interaction becomes persistent, users may benefit from personalization, but lose awareness of how data shapes incentives and monetization strategies. This tension is amplified by AI systems that increasingly infer player traits, predict behavior, and adapt mechanics without transparent explanation. At the same time, users have limited means to understand or contest such processes, which heightens concerns about autonomy and fairness. Meanwhile, gamification has been shown to cause users to be more likely to share personal information (Trang & Weiger (2021)), which means that data privacy has to be analysed differently in this context. Gamification is defined through the presence of ludic elements in non-ludic contexts, such as points, badges, and leaderbords. Research on the topic is growing rapidly and shows the need for more critical engagement with the topic (Tobon et al. (2020)). Privacy concerns in gamified apps have been studied in detail before (Mavroeidi et al. (2019); Kapoor (2025)), but have not connected these issues with the growing concerns about generative artificial intelligence. Similarly, the perception of specific gamification elements has been studied (Garcia-Iruela & Hijon-Neira (2020)), but not linked to issues of data privacy.

These developments illustrate the need for empirical insight into how players interpret the risks and benefits of modern gamification. However, it remains unclear how players themselves perceive these developments, especially regarding privacy, manipulation risks, and the role of regulation. This gap motivates our empirical study.

The main contributions of this work are:

- We investigate player awareness of data collection and AI-driven inference in games and gamified applications.
- We analyze how players evaluate leveling systems, leaderboards, daily login rewards, and random reward reveals with respect to motivation and manipulation.
- We examine player expectations toward regulatory safeguards against exploitative design.

This study provides initial empirical evidence on how users interpret the goals and risks associated with contemporary gamification features. Section 2 describes how modern ergonomics technologies enable the development of gamification solutions. Section 3 discusses the dual dynamics of innovation and ethical risk. Section 4 outlines the study design. Section 5 presents empirical findings, and Section 6 summarizes implications for responsible, regulated gaming.

## 2. CONVERGENCE OF GAMING, AI, AND EMERGING COMMUNICATION TECHNOLOGIES

Computing environments are increasingly shaped by the demand for interactive experiences with strict latency guarantees and data-driven intelligence (Horvath et al. (2025)). Modern gaming is no longer bound to a single platform but distributed across smartphones, wearable devices, AR/VR headsets, and cloud-enabled consoles (Hamadanian et al. (2023); Yu et al. (2020)). Players expect continuity across devices and contexts, which pushes system designers to support seamless cross-platform interaction and consistent performance. Location-based gameplay and sensor-driven mechanics exemplify how digital play extends into everyday environments (Hamari et al. (2019)). AI has become a central driver of this evolution. In games, AI supports adaptive difficulty, personalized content delivery, and autonomous agents (Tucek (2024); Yannakakis and Togelius (2018)). Beyond the game itself, edge and cloud intelligence enable dynamic orchestration of computational tasks, reducing device load while maintaining responsiveness (Horvath et al. (2025)). These capabilities blur the boundaries between local interaction and distributed computation, allowing for experiences that react to user context in real-time.

## 3. CHALLENGES AND OPPORTUNITIES

The convergence of gaming, AI, and emerging communication technologies creates an aforementioned dual dynamic, with technical systems functioning as enablers for pervasive, data-driven play, and societal systems lacking adequate safeguards to protect users. On the technological side, latency-sensitive architectures, cross-device interaction, and adaptive AI models expand the possibilities of what gaming experiences can offer. On the societal side, these same capabilities introduce forms of behavioral influence, personal data extraction, and economic exploitation that are not yet sufficiently regulated or well understood. The following subsections outline how these challenges emerge from technological enablers and how the absence of regulatory clarity affects individuals and society.

## 3.1 Technical Potential and Behavioral Risk

Modern game experiences rely on seamless responsiveness and continuity across devices. Technologies such as edge computing, distributed inference, and adaptive networking enable real-time interaction and personalized challenges (Hamadanian et al. (2023); Horvath et al. (2025); Yannakakis and Togelius (2018)). These capabilities support scalable and immersive play even on resource-constrained mobile or wearable devices (Entertainment Software Association (2024)). As AI systems observe skill level, preferences, and play habits, they can tailor difficulty and content to foster engagement and a feeling of progress.

At the same time, these design choices introduce behavioral and ethical risks. Progress metrics, ranking systems, and reward schedules leverage psychological reinforcement (Kim and Castelli (2021)) and can pressure users into habitual interaction rather than intentional play. When telemetry informs adaptive incentives or monetization, players become targets of commercial optimization rather than active participants, raising concerns about fairness, autonomy, and emotional well-being (Mavroeidi et al. (2019); Peters (2020)). Users often lack visibility into how personal data shapes the mechanics that retain their attention, and opting out may require abandoning the experience entirely. These tensions motivate the need to understand how players themselves interpret the balance between motivation and manipulation in contemporary gamification.

## 3.2 Comparison between Technology, Governance, and Societal Implications

The technical enablers that support gaming, low-latency networking, cross-device continuity, and AI-driven personalization operate at a speed that exceeds the response capabilities of regulatory systems. Existing frameworks target consumer protection in narrow domains, such as data ownership or platform monetization, but fail to address the behavioral dynamics introduced by adaptive systems. Without transparent safeguards, individual users face information asymmetry, weak consent boundaries, and subtle manipulation of engagement. At the societal level, these practices normalize constant performance tracking, reward-driven competition, and commercial extraction of everyday behavior.

## 3.3 Research Questions

Advances in personalization and engagement design have transformed how players interact with contemporary games and gamified applications. However, these same mechanisms introduce uncertainty about data usage, behavioral influence, and the adequacy of existing oversight. To investigate these tensions, we focus on three research questions:

> RQ1: How concerned are players about personal-data collection and AI-driven inference in games and gamified applications?

RQ1 examines the growing tension between personalization and surveillance. Games routinely collect telemetry for matchmaking, progression, and targeted incentives. Yet users are rarely informed about how these data shape experience. We investigate whether players view such practices as beneficial personalization, unclear commercial extraction, or a threat to autonomy and digital dignity (Becker (2021)).

> RQ2: Which gamification features do players experience as motivating, and which are manipulative or coercive?

RQ2 focuses on how design choices affect player motivation and activity. Mechanics such as leveling systems and leaderboards can foster progress and mastery, while retention-driven strategies may pressure players to engage customarily. By eliciting player perceptions, we identify where gamification supports competence and where it risks creating compulsive or exploitative interactions.

> RQ3: How aware are players of existing protections against exploitative design, and what regulatory action do they expect?

RQ3 explores regulatory awareness and expectations. Users frequently assume that platform policies or laws prohibit harmful design patterns, yet many such protection measures do not meaningfully address behavioral manipulation. We investigate whether players feel sufficiently protected, and to what extent they call for stronger intervention to safeguard vulnerable users and ensure accountability in gamified systems.

# 4. USER STUDY

Although limited in scope and exploratory in nature, our initial study provides a detailed understanding of how players perceive gamification systems, including their concerns about data and AI usage, their assessment of motivational design elements, and their expectations regarding regulatory oversight.

## 4.1 Survey Instrument

We designed an anonymous online questionnaire with twelve items. The survey consisted of three sections. The first section included informed consent and age eligibility requirements. The second section collected demographic information (age and weekly gaming hours) and included six items on privacy concerns, awareness of data usage, perceived control of gameplay data, and attention to developments in artificial intelligence. These items used 7-point Likert scales (Likert (1932)) with anchors from Strongly disagree (1) to Strongly agree (7). These items are taken from established information privacy instruments, with the first three stemming from the IUIPC scale (Malhotra et al. (2004); Groß (2021)) and the second half from PC-AIM (Menard & Bott (2024)). We adapted the items to focus specifically on games and gamified apps.

The final section asked participants to rate four gamification features: leveling systems, leaderboards, daily login rewards, and random reward reveals. Two optional open-ended questions invited respondents to describe privacy concerns or positive experiences with gamification.

## 4.2 Procedure

The survey was created using Google Forms and was distributed through university channels and gaming-focused online communities. Participation was voluntary and anonymous, and only adults aged 18 or older were eligible. No personally identifying data was required. Before beginning the survey, participants confirmed that they consented to the use of their anonymized responses for academic research. The analysis includes statistics for all quantitative items, such as mean scores and response distributions. Qualitative responses will be evaluated through thematic grouping in order to identify recurring perspectives or concerns.

# 5. RESULTS

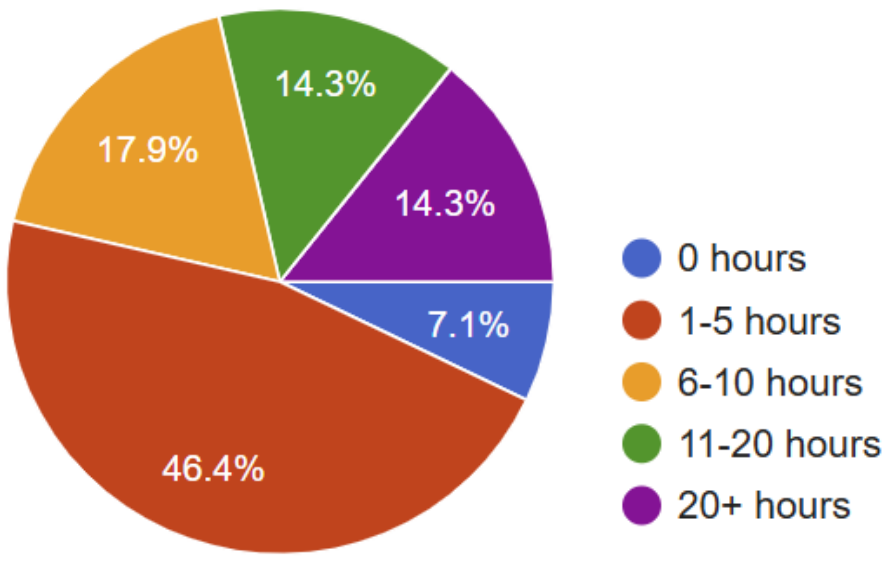

Figure 1 Weekly gaming-time distribution across the 28 participants.

The survey provides an empirical perspective on how players interpret data practices, gamification features, and regulatory safeguards in gaming environments. We analyzed responses from N = 28 adult participants, all of whom confirmed their eligibility and consent. Despite the small sample size, the represented demographic's age range is 21 to 41 years (M = 29.4, SD = 6.0), and reflects a diversity of gaming habits.

Figure 1 depicts the weekly gaming time: 7.1% do not play regularly, 46.4% play 1-5 hours per week, 17.9% play 6-10 hours, 14.3 % play 11-20 hours, and 14.3 % play for more than 20 hours each week. This distribution, indicates that our sample includes both casual and highly engaged players, providing a useful spectrum of perspectives for evaluating motivation, manipulation, and attitudes toward privacy and AI.

## 5.1 Privacy and AI Concerns (RQ1)

Participants expressed high concern regarding personal data collection and AI-driven inference. In contrast, awareness of existing practices and perceived control over data were notably lower, indicating an information-gap risk. Table 1 summarizes these outcomes.

Although many participants expressed concerns about personal data collection (mean = 5.44 on a 7-point scale), far fewer felt that they had meaningful control over their data (mean = 2.56). This contrast indicates a lack of clarity from games and gamified apps regarding how users can manage or limit data processing.

Table 1 Privacy and AI Concern Ratings (1–7 Likert Scale)

| *Item* | *Mean* | *SD* | *Median* |
|---|---|---|---|
| Concern about data collected | 5.44 | 1.36 | 6 |
| Awareness of data usage | 3.64 | 1.54 | 4 |
| Control over data usage | 2.56 | 1.39 | 2 |
| Concern about AI misuse | 5.64 | 1.32 | 5 |
| Following AI developments | 4.36 | 1.78 | 4 |
| Concern about AI developments | 5.32 | 1.51 | 6 |

A similar pattern appears in attitudes toward AI developments (mean = 5.32). Because players cannot easily access or adjust the data collected about them, they also lack visibility into how these data inform AI-driven features or contribute to model training. Open-text responses reinforced these concerns, frequently citing targeted advertising, profiling, and exploitation of vulnerable users. Several participants worried AI may enable non-transparent decisions or normalize surveillance practices. These findings align with societal challenges described in Section 3.2.

## 5.2 Motivation vs. Manipulation of Gamification (RQ2)

Participants drew a clear distinction between constructive progression mechanics and retention-oriented reward systems. As shown in Table 2 and visualized in Figures 2 and 3, leveling systems were overwhelmingly viewed as motivating, with most responses clustering at the upper end of the scale. In contrast, leaderboards, daily login rewards, and random reward reveals elicited far more neutral or low motivational judgments. Leaderboards displayed a broad and relatively flat distribution, indicating that while some players find them engaging, many do not consider them a major motivational factor. Random reward reveals also received predominantly low motivation ratings, suggesting that surprise-based rewards do not foster sustained engagement for most respondents.

Table 2 Motivation vs. Manipulation Ratings (1–7 Likert Scale)

| *Feature* | *Motivation Mean* | *Manipulation Mean* | *Gap* | *Interpretation* |
|---|---|---|---|---|
| Leveling systems | 5.20 | 3.72 | +1.48 | Rewarding progress |
| Leaderboards | 3.84 | 4.84 | −1.00 | Competitive pressure |
| Daily login rewards | 2.60 | 5.60 | −3.00 | Habit enforcement |
| Random reward reveals | 3.00 | 6.24 | −3.24 | Gambling-like |

Perceptions of manipulation reveal a reciprocal pattern. Leveling systems are not only rated as highly motivating but are also considered the least manipulative feature, with responses concentrated near the lower end of the manipulation scale. This rare combination of high motivation and low perceived manipulation helps explain why leveling mechanics remain widely accepted in both games and gamified applications.

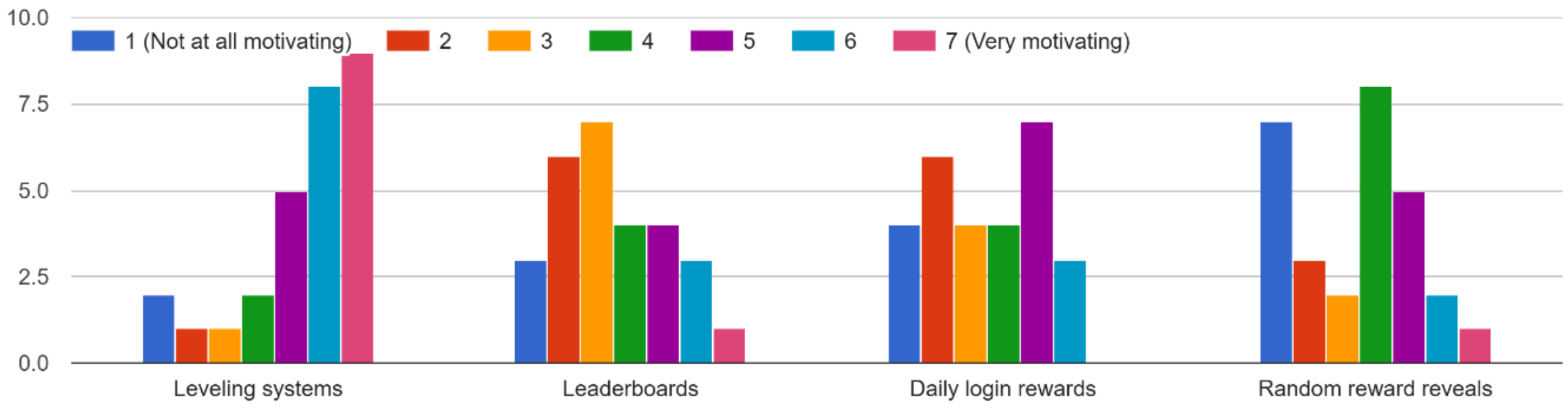

Figure 2 Distribution of motivation ratings (1-7) for the four evaluated gamification features

Random reward reveals show the most pronounced opposite trend. Manipulation ratings focus heavily toward the upper end of the scale, indicating that participants perceive these systems as highly manipulative despite finding them minimally motivating. Meanwhile, leaderboards occupy an intermediate position: although they offer only modest motivational value, players do not uniformly interpret them as manipulative. Daily login rewards, however, present a clearer pattern of perceived manipulation, reinforcing concerns about habit-enforcement loops and fear-of-missing-out mechanics.

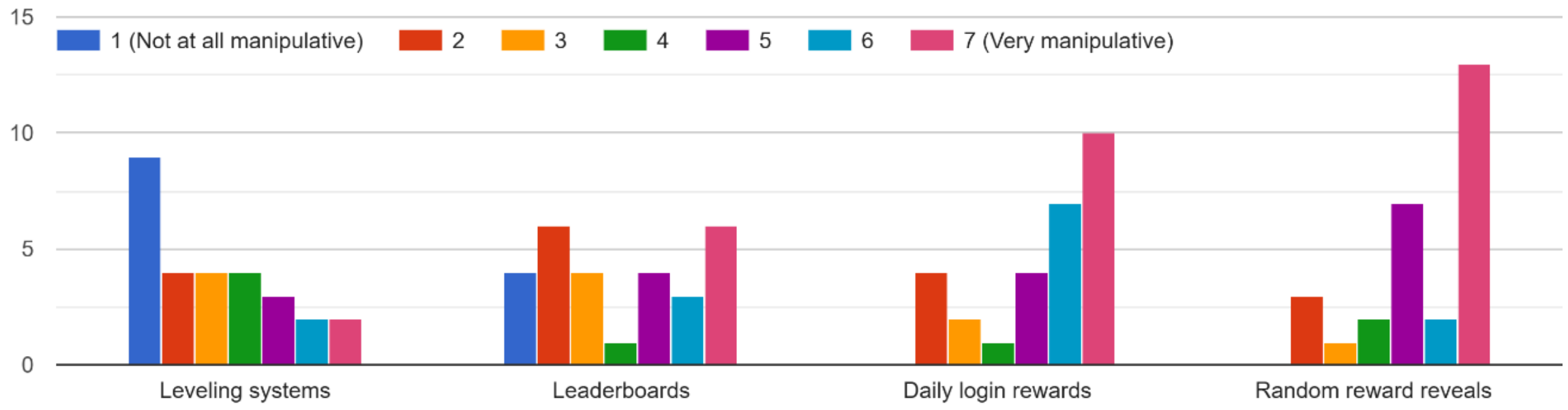

Figure 3 Distribution of manipulation ratings (1-7) across the same features.

Qualitative comments support this overall interpretation. Several participants described tools such as Duolingo streaks or seasonal battle passes as simultaneously motivating and coercive. This dual perception highlights how gamification mechanics can promote engagement while also generating pressure or dependency, particularly among users who may be more susceptible to sustained retention strategies.

## 5.3 Regulatory Expectations (RQ3)

Participants showed substantial uncertainty about what regulations apply to games and gamified applications. Many indicated that they were unsure whether any protective measures exist, or how data use, AI-driven profiling, or monetization strategies are currently governed. This lack of awareness mirrors the patterns observed in RQ1: users express strong concern about personal data practices and AI, yet they do not feel informed enough to understand what safeguards are in place or how they might assert their rights.

Open-text responses reinforced this uncertainty and highlighted worries about how manipulative mechanics may affect vulnerable populations. Several participants noted that children, elderly users, or individuals with cognitive challenges may be especially susceptible to pressure-based design and may more easily disclose personal information or spend money unintentionally. Respondents also emphasized that manipulative reward systems tend to be demotivating rather than engaging, a pattern clearly reflected in the questionnaire results shown in Figures 2 and 3. Participants frequently described such mechanics as both ethically concerning and counterproductive, underscoring a desire for clearer standards that limit psychological pressure while preserving genuinely motivating forms of play.

# 6. CONCLUSION

This paper examined how gaming and AI-driven personalization reshape both the technical foundations and societal implications of play. While network-native experiences benefit from low-latency architectures, distributed computation, and adaptive mechanics, our findings reveal that these advances coexist with significant concerns about data use, behavioral influence, and the transparency of modern gamification systems. Across 28 participants, we observed a consistent gap between strong privacy concerns and limited perceived awareness or control. Users recognize that games and gamified applications collect extensive behavioral data, yet they feel poorly informed about how this data is transformed into AI-driven inference, targeted incentives, or monetization strategies. This mismatch between concern and agency highlights a structural problem: players are embedded in data-intensive systems without the means to understand or meaningfully negotiate their role within them.

Our results further show that players differentiate sharply between motivating and manipulative gamification mechanics. Progress-oriented features such as leveling systems are widely appreciated because they signal mastery and sustained effort while preserving user autonomy. Leaderboards occupy a more ambivalent position, motivating some players while discouraging others, depending on competitive preference and social context. In contrast, retention-focused mechanisms including daily login rewards and random reward reveals are overwhelmingly perceived as manipulative. The distributional analyses in Figures 2 and 3 demonstrate a clear alignment: features that players view as manipulative are simultaneously demotivating, suggesting that coercive strategies undermine rather than enhance meaningful engagement. These mechanics are often associated with fear-of-missing-out, pressure to maintain streaks, or gambling-like reward structures, which several participants identified as especially harmful for children or other vulnerable users.

Participants also expressed uncertainty about what regulatory protections exist and whether current frameworks meaningfully limit exploitative design. Open-text responses emphasized that users lack clarity about how data is governed, how AI is trained, and what safeguards constrain monetization practices. Many respondents voiced concern that minors, elderly users, or individuals with cognitive challenges may be particularly exposed to manipulative mechanics.

The findings of this study align with prior research (Garcia-Iruela and Hijón-Neira (2020)), who reported that progression-oriented gamification elements such as levels and points are perceived more positively than competitive mechanics like leaderboards. A similar pattern emerges here, with leveling systems viewed as motivating and less manipulative, while competitive and retention-focused features are evaluated more critically. Taken together, these results indicate that technical innovation alone is insufficient for responsible digital play; as games and gamified applications increasingly permeate everyday activities, designers and platform operators must align engagement strategies with user well-being rather than attention extraction. This requires transparency mechanisms that make data flows and AI-driven adaptations understandable to non-experts, as well as meaningful consent options that allow users to opt out of data-driven personalization or disengage without penalty. Ethical design should therefore prioritize user autonomy, reduce pressure-based feedback loops, and reinforce competence rather than compulsion.

This study provides empirical evidence that users clearly differentiate between motivating and manipulative gamification mechanics. We extended prior work by linking these perceptions to data-driven personalization and AI-enabled behavioral influence, underscoring the need for transparent and ethically aligned design.

Future research should explore how players interpret explainable AI in gaming contexts, how behavioral data can be governed more democratically, and how regulatory frameworks can better address manipulative design without stifling innovation. A socio-technical approach that integrates engineering with accountable governance is necessary to ensure that pervasive gamification remains a benefit to players, rather than a risk to their autonomy, privacy, and wellbeing. By grounding design decisions in fairness, transparency, and user empowerment, the industry can move toward game experiences that motivate without manipulating and personalize without exploiting.